\documentclass[10pt,conference]{IEEEtran} 
\IEEEoverridecommandlockouts
\usepackage{cite}
\usepackage{amsmath,amssymb,amsfonts}
\usepackage{algorithm}
\usepackage{algpseudocode} 
\usepackage{makecell}
\usepackage{graphicx}
\usepackage{multirow}
\usepackage{textcomp}
\usepackage{xcolor}
\usepackage{needspace}
\usepackage{float}
\usepackage{hyperref}

\def\BibTeX{{\rm B\kern-.05em{\sc i\kern-.025em b}\kern-.08em
    T\kern-.1667em\lower.7ex\hbox{E}\kern-.125emX}}

\begin{document}

\title{A First Step Towards Detecting Values-violating Defects in Android APIs}

\author{\IEEEauthorblockN{Conghui Li}
 \IEEEauthorblockA{\textit{Faculty of Information Technology} \\
 \textit{Monash University}\\
 Melbourne, Australia \\
 clii0087@student.monash.edu}
 \and
 \IEEEauthorblockN{Humphrey O. Obie}
 \IEEEauthorblockA{\textit{HumaniSE Lab} \\
 \textit{Monash University}\\
 Melbourne, Australia \\
 humphrey.obie@monash.edu}
 \and
 \IEEEauthorblockN{Hourieh Khalajzadeh}
 \IEEEauthorblockA{\textit{HumaniSE Lab} \\
 \textit{Monash University}\\
 Melbourne, Australia \\
 Hourieh.Khalajzadeh@monash.edu}}

\maketitle

 \begin{abstract}
Human values are an important aspect of life and should be supported in ubiquitous technologies such as mobile applications (apps). There has been a lot of focus on fixing certain kinds of violation of human values, especially privacy, accessibility, and security while other values such as pleasure, tradition, and humility have received little focus. In this paper, we investigate the relationship between human values and Android API services and developed algorithms to detect potential violation of these values. We evaluated our algorithms with a manually curated ground truthset resulting in a high performance, and applied the algorithms to 10,000 apps. Our results show a correlation between violation of values and the presence of viruses. Our results also show that apps with the lowest number of installations contain more violation of values and the frequency of the violation of values was highest in social apps.
\end{abstract}


\section{Introduction}
Mobile applications (apps) are widely used in the modern society to improve our productivity and to fulfil end-users purposes \cite{b2}. However, the use of technology can sometimes violate the values of end-users and in the worst case cause harm to them and society at large. For example, a recently publicised employment interview algorithms were shown to be flawed, when they scored a reporter 6 out of 9 points for English-language competency after the interview was conducted exclusively in German \cite{Bloomberg:2021}. Such a flawed system could bar qualified candidates from getting jobs and cause employees to miss out on good candidates.

Human values are the guiding principles for what people consider important in life \cite{b8}. They serve as a guide for actions and a vehicle for expressing need.
Human values such as privacy and security have garnered both academia and industry focus but other broader values including self-direction, benevolence, tradition, etc. have not received sufficient attention in the research community \cite{Perera:2020}. Although there have been some recent efforts in value-based software engineering, the field is still in its nascent stages \cite{Perera:2020}.


One of the earliest attempts in describing human values at a low level is the recent work by Mougouei\cite{b1}, which proposed a theoretical framework named, \textit{AIR} (Annotation, Inspection, and Recommendation) to account for human values in source code. Specifically, this framework includes the following components: annotating Application Programming Interfaces (APIs) with human values based on the relationship between human values definition and the API definition; annotating code at class level or method level based on the identified values in the APIs; using the results to inspect values that may be potentially breached by the code, leading to a condition named \textit{value smell}; and finally, recommending a possible fix for the potential values violation. Our work builds upon and extends this theoretical framework by concretising the detection of potential violation of values in source code.


In this paper, we investigate the relevant relationships between common API services and human values based on the literature and introduce multiple algorithms based on the Android API services to detect potential violation of values in Android apps. We evaluate our algorithms on an evaluation set; a manually annotated source code from 46 android apps with an average accuracy of 0.93 and recall of 0.84. Prompted by the promising evaluation results, we applied our models to 10,000 apps randomly curated from AndroZoo \cite{b11}, covering multiple categories. Our results show a correlation between violation of values and the presence of viruses. Our results also show a correlation between the violation of human values and the presence of viruses in apps; apps with the lowest number of installations had more potential of value violations; and social apps had the highest rate of value violation.




\section{Background and Motivation}
Human values determine behaviour and attitude. \cite{Cheng:2010}. The concept of human values have been well explored and developed in the social sciences.

Human values play a critical role in our lives to explain the motivations, opinion types and actions for different cultural groups, and individuals \cite{b8}. The Schwartz theory of basic human values is the most widely accepted and adopted model of human values. This theory is grounded on surveys conducted in multiple countries and covers a wide range of different ages, cultural backgrounds, genders, and geography \cite{b8}. The Schwartz categorises 58 value items in 12 broad categories. These categories are described in \autoref{table:Schwartz theory}. Schwartz theory seen adoption in different areas including software engineering, e.g., the Values Q-sort has been used to understand the values opinion-types of software engineers \cite{Winter:2018}. A similar work adopted the Schwartz theory to understand the values of Bangladeshi female farmers’ for agriculture mobile applications development \cite{Shams:2020}. Recently, Obie et al. developed natural language processing (NLP) techniques to understand potential violation of human values in app reviews as reported by users \cite{Obie:2020}.

Mougouei argues that APIs such as those implemented in Android source code are not values-agnostic, and introduced the AIR framework as an approach to support value programming in software projects \cite{b1}. Mougouie proposed relationships between 9 common Android apps and human value categories (c.f. \cite{b1}). Based on these relationships, APIs can violate certain human values, e.g., the values of security and universalism can be violated by the Android Accessibility Service if not properly implemented \cite{b3}. Human values can be violated either through negligence or through deliberate abuse of API services. While \cite{b1} proposes a framework, our work is the first to design algorithms for detecting potential value-violations and conducting a large scale study to understand the prevalence of these violations in Android apps.

\begin{table}
\scriptsize
\centering
\caption{Value categories and description \cite{b8}}
\label{table:Schwartz theory}
\begin{tabular}{lp{6cm}} 
\hline
\textbf{\textit{Value Category}} & \multicolumn{1}{c}{\textbf{\textit{Description}}}                                                                         \\ 
\hline
Self Direction                   & Independent thought and action—choosing,creating, exploring                                                               \\ 
\hline
Stimulation                      & Excitement, novelty, and challenge in life                                                                                \\ 
\hline
Hedonism                         & Pleasure and sensuous gratification for oneself.                                                                          \\ 
\hline
Achievement                      & Personal success according to social standards                                                                            \\ 
\hline
Power                            & Social status and prestige, control or dominance over people and resources                                                \\ 
\hline
Security                         & Safety, harmony and stability of society, of relationships, and of self                                                   \\ 
\hline
Tradition                        & Respect, commitment and acceptance of the customs and ideas that traditional culture or religion provide                  \\ 
\hline
Conformity                       & Restraint of actions, inclinations, and impulses likely to upset or harm others and violate social expectations or norms  \\ 
\hline
Benevolence                      & Preservation and enhancement of the welfare of people with whom one is in frequent personal contact                       \\ 
\hline
Universalism                     & Understanding, appreciation, tolerance and protection for the welfare of all people and for nature                        \\ 
\hline
Face                             & maintaining public image and avoiding humiliation                                                                         \\ 
\hline
Humility                         & Recognizing insignificance in the larger scheme of things                                                                 \\
\hline
\end{tabular}
\end{table}

\begin{table*}
\scriptsize
\centering
\caption{Relationship between selected APIs and human values. Adapted from Mougouei \cite{b1}}
\label{table:API violation}
\begin{tabular}{lp{6cm}llllll} 
\hline
\textbf{\textit{API name}} & \textbf{API description}                                                                                                          & \textbf{Self-direction} & \textbf{Stimulation} & \textbf{Hedonism} & \textbf{Face} & \textbf{Tradition} & \textbf{Humility}  \\ 
\hline
Android.animation          & These classes provide functionality for the property animation~system, which allows you to animate object properties of any type. &                         &                      & \checkmark               &               &                    &               \\ 
\hline
Android.media              & Provides classes that manage various media interfaces in audio~and video.                                                         & \checkmark                     & \checkmark                  & \checkmark               &               &                    & \checkmark           \\ 
\hline
Android.mtp                & Provides APIs that let you interact directly with connected cameras and other devices, using the Picture Transfer Protocol (PTP)~ & \checkmark                 &                      &                   &               &                    &               \\ 
\hline
Android.nfc                & Provides access to Near Field Communication (NFC) functionality,~allowing applications to read NDEF message in NFC tags.          & \checkmark                     &                      &                   &               & \checkmark                &               \\ 
\hline
Android.telephony          & Provides APIs for monitoring the basic phone information, plus~utilities for manipulating phone number strings.                   &                         &                      &                   & \checkmark           & \checkmark                &               \\ 
\hline
Android.hardware           & Provides support for hardware features, such as the camera and~other sensors.                                                     &                         &                      &                   & \checkmark           &                    &               \\
\hline
\end{tabular}
\end{table*}


\begin{table*}
\scriptsize
\centering
\caption{Overview of evaluation truthset data}
\label{table:API violation of categories in experiment}
\begin{tabular}{|l|l|l|l|l|l|l|l|l|l|l|l|} 
\hline
Human value                     & API name  & communication & finance & education & fitness & game & health & photography & productivity & social & travel  \\ 
\hline
\multirow{3}{*}{Self Direction} & mtp       & 0             & 0       & 0         & 0       & 0    & 0      & 0           & 0            & 0      & 0       \\ 
\cline{2-12}
                                & media     & 9             & 5       & 3         & 3       & 2    & 1      & 1           & 2            & 0      & 0       \\ 
\cline{2-12}
                                & nfc       & 0             & 0       & 0         & 0       & 0    & 1      & 0           & 0            & 0      & 0       \\ 
\hline
\multirow{3}{*}{Security}       & hardware  & 0             & 0       & 0         & 0       & 0    & 0      & 0           & 0            & 0      & 0       \\ 
\cline{2-12}
                                & telephony & 3             & 0       & 1         & 0       & 0    & 0      & 0           & 0            & 0      & 0       \\ 
\cline{2-12}
                                & nfc       & 1             & 0       & 0         & 0       & 0    & 0      & 0           & 0            & 0      & 0       \\ 
\hline
\multirow{2}{*}{hedonism}       & media     & 0             & 4       & 3         & 0       & 2    & 0      & 1           & 1            & 0      & 0       \\ 
\cline{2-12}
                                & animation & 0             & 0       & 0         & 0       & 0    & 0      & 0           & 0            & 0      & 0       \\ 
\hline
Universalism                    & media     & 0             & 4       & 3         & 0       & 2    & 0      & 1           & 1            & 0      & 0       \\ 
\hline
Conformity                      & telephony & 3             & 0       & 1         & 0       & 0    & 0      & 0           & 0            & 0      & 0       \\ 
\hline
Total number                    &           & 14            & 7       & 7         & 7       & 2    & 3      & 1           & 3            & 1      & 1       \\
\hline
\end{tabular}
\end{table*}


\section{Rule-based Values Violation Detectors}
In this paper, we selected a subset of 6 Android APIs related to various human values. Our selection is based primarily on API usage frequency as documented by Wang and Godfrey \cite{b10}. Using the Android documentation\footnote{https://developer.android.com/reference/packages}, we also selected some APIs that support hardware interaction and user management. These include the following: \textit{android.animation}, \textit{android.media}, \textit{android.mtp}, \textit{android.nfc}, \textit{android.telephony} and \textit{android.hardware}. We show the selected API's and their related values in \autoref{table:API violation}. We developed algorithms to detect potential violation of values in the use of selected APIs in Android apps.

\subsection{Data Collection}
We collected 46 Android apps from the Google Play store covering 10 categories and then manually annotated them with value violation -- API tags. The first author did a round of annotation which was validated by the second author, and based on discussions during Zoom meetings between the first and second author, the annotations were finalised. \autoref{table:API violation of categories in experiment} shows the application categories, APIs, and number of potential value violations. These annotated apps formed our truthset used for the evaluation of our algorithms.

\subsection{Data Preparation}
The preparation of our data for the detection of values-violation in the selected API's is done in two simple phases: Android Application Package (APK) decomposition and Java Abstract Syntax Tree (AST) analysis.

\subsubsection{APK Decomposition} the source in APK is sealed by android, hence it needs to be decompiled to be able to access the source code. We used the APKtool \footnote{https://ibotpeaches.github.io/Apktool/} in this phase.
    
\subsubsection{AST Analysis} typically, a Java program without bug can be represented in an abstract syntax tree structure, and thus it is possible to traverse the tree. We traverse the AST using Depth-First Search and identify code that may be linked to values violation. An example detector for the violation of Universalism value in Media API is shown in Algorithm. \autoref{algorithm:Media universalism}.

\begin{algorithm}
\footnotesize
	\caption{Media API/Universalism value violation detector}
    \label{algorithm:Media universalism}
	\begin{flushleft}
        \textbf{INPUT:} : $JavaPath$ java file path\\
        \textbf{OUTPUT:} return a boolean value: \\
            \qquad $True$: violation exist \\
            \qquad $False$: violation does not exist
    \end{flushleft}
	\begin{algorithmic}[1]
	    \State /*using deep first search to search all the AST nodes out*/
	    \State $AstNodes \leftarrow $ search( $JavaPath$ )
		\State $ViolationList\leftarrow $ an empty list
		\State $VarNames\leftarrow $ an empty list
		\For {\textbf{all} $AstNode$ $in$ $AstNodes$}
		\If{$AstNode$ $is$ Import Decelerator Node}
		\If{\textbf{``com.google.android.exoplayer"} = $AstNode$.path}
		\State append \textbf{"Violation"} $in$ $ViolationList$
		\EndIf
		\EndIf
		\EndFor
		
		\For{ \textbf{all} $ListElement$ $in$ $ViolationList$}
		\If{$ListElement$ = \textbf{``Violation"}}
		\State \textbf{Return} True
		\EndIf
		\EndFor
		\State \textbf{Return} False
		
	\end{algorithmic} 
\end{algorithm}

\subsection{Values Violation Detection Methods}
We introduce an overview of the APIs and algorithms for detecting potential violation of values. The complete set of algorithms and source codes are provided in a supplementary repository\footnote{https://github.com/awesomehumphrey/Values-violating-defects-in-android-apis}.
\subsubsection{Media} 
The media API\footnote{https://developer.android.com/reference/android/media/package-summary} provides the functionality of multimedia including audio, video, and manage media data for users. Based on \cite{b1}, using the media API can potentially violate human values. 

Firstly, multimedia advertisements (ads) can lead to the violation of self-direction, especially when the ads are forced on the users without an option to close the ad. Also, in cases where the presented ads lack diversity, representing the different background of users, this can constitute a violation of the value of universalism. To detect this potential violation, our algorithm flags import statements related to advertisements, as shown in Algorithm \autoref{algorithm:Media universalism}.

Secondly, automatic background music without controls for stopping or pausing might violate the values of pleasure and independence of a user. To detect this, we check the code combination where the \textit{MediaPlayer} object only provides a function for playing media without a stop or pause control. 






\subsubsection{Animation}  
The animation API\footnote{https://developer.android.com/reference/android/animation/package-summary} applies to the animation system in android, and is used to animate properties of any kind. The animation API is related to the value of hedonism. A good animation could help users understand an application logic, increase the usability of an app, and bring pleasure to a user. However, animation when done badly may negatively affect the user's experience of using an app and potentially breach their value of pleasure, e.g. when an animation is excessively slow or endless due to an infinite loop. To detect these potential violations in the animation API, we check to see if the parameter of the \textit{setRepeatCount()} method is set to infinity and/or if the duration time in the \textit{SetDuration()} method exceeds two seconds.


\subsubsection{MTP} 
The MTP API\footnote{https://developer.android.com/reference/android/mtp/package-summary} provides the functionality of interaction among user's device, camera and other devices with the Picture Transfer Protocol (PTP). The MTP API is related to the value of self-direction (i.e., independence). To detect potential violation of the value of self-direction, our algorithm checks to see whether the data transmission is bi-directional;  if the \textit{MtpDevice} objects do not call the input and output data methods in the same class, it is flagged as a potential value violation.


\subsubsection{NFC}
\begin{figure}[!htb]
\includegraphics[width=0.5\textwidth]{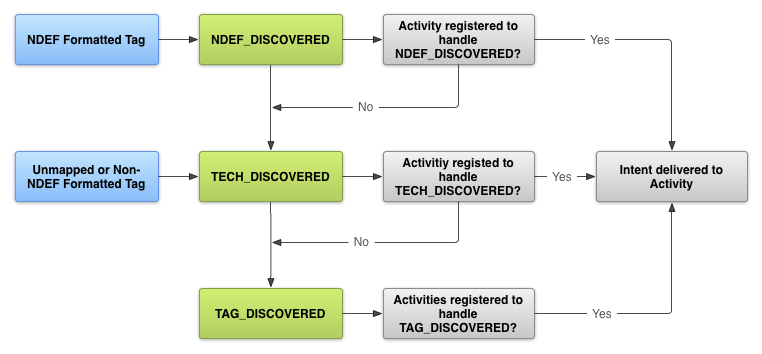}
\caption{NFC Tag Dispatch System}
\centering
\label{fig:nfc system}
\end{figure}

Near Field Communication (NFC)\footnote{https://developer.android.com/reference/android/nfc/package-summary} is a technique widely applied in our daily life. For instance, with the NFC users' phones can be used as digital wallets.
The NFC tag consists of 2 types of the data format:  NFC Data Exchange Format (NDEF) or Non-NDEF. Moreover, to distribute the NFC tag data to the correct activities precisely, the tag dispatch system provides 3 NFC action intents, including \textit{NDEFDISCOVERED}, \textit{TECH DISCOVERED} and \textit{TAG DISCOVERED}, for different distribution priority. The priority structure is shown \autoref{fig:nfc system}. Also, the intent \textit{NDEFDISCOVERED} can only be manipulated by NDEF data.


Typically, a well-operated system only distributes the required resource or authority to an activity. As such, our algorithm flags the potential violation of the value of self-direction if the developer does not set the intents in the order of intent priority in the hierarchical structure, e.g., if the NDEF tag is first distributed to the \textit{TAG DISCOVERED} instead of \textit{NDEFDISCOVERED}. After inspecting all intents in the manifest file, the violation of the self-direction value would be flagged  if intent with low priority supercedes intent with higher priority.

In general, when applications write information into NFC tags, developers may need to ensure that the tags have security mechanisms to prevent issues like privacy leaks. One effective solution is to insert the Android Application Records (AAR). To check for a potential violation of privacy, our algoritm checks whether the  method call \textit{createApplicationRecord()} from  \textit{NdefRecord} object exists.



\subsubsection{Telephony}
The telephony API\footnote{https://developer.android.com/reference/android/telephony/package-summary} provides essential information and status of cell phone, and also serves as the communication monitor. Because the function of sending an SMS message to other users through their phone numbers is prone to eaves-dropping and possible manipulation, there exist the potential violation of the values of security and conformity. Taking the method \textit{SendMultipleMessage()} as an example, sending and being called by \textit{SmsManager} object will be queried over all java files of the application by our algorithm.


\subsubsection{Hardware}  
The hardware API\footnote{https://developer.android.com/reference/android/hardware/package-summary} support developers in manipulating sensors, camera, and other components in devices. The human value associated with the hardware API is privacy and security. Using hardware API, the most significant privacy and security issue is that an app can collect photographs or other media data without users' consent. Since photographs are one of the most important private and personal identifying information of users\footnote{https://www.oaic.gov.au/privacy/guidance-and-advice/mobile-privacy-a-better-practice-guide-for-mobile-app-developers/}, our algorithm detects automatic photograph capture and invisible camera layout. Our algorithm tracks the method \textit{takePicture()} called by Camera object and \textit{createCaptureSession()} called by \textit{CameraDevice object}, as well as the surface layouts with a small size.



%
%
%
%


\begin{table*}[!h]
\scriptsize
\centering
\caption{frequency of violation of values for different App categories}
\label{table:violation frequency for categories}
\begin{tabular}{|l|c|l|l|c|c|c|c|c|} 
\hline
app categories  & count & violation & rate   & Hedonism & Self Direction & Universalism & Security & Conformity  \\ 
\hline
ART             & 17    & 1         & 5.8\%  & 0        & 1              & 1            & 0        & 0           \\ 
\hline
AUTO            & 31    & 2         & 6.4\%  & 0        & 2              & 2            & 0        & 0           \\ 
\hline
BEAUTY          & 17    & 1         & 5.8\%  & 0        & 0              & 0            & 1        & 1           \\ 
\hline
BOOKS           & 177   & 12        & 6.7\%  & 1        & 11             & 10           & 1        & 0           \\ 
\hline
BUSINESS        & 262   & 22        & 8.3\%  & 1        & 18             & 17           & 5        & 3           \\ 
\hline
COMICS          & 7     & 1         & 14.3\% & 0        & 0              & 0            & 1        & 1           \\ 
\hline
COMMUNICATION   & 80    & 9         & 11.2\% & 0        & 8              & 7            & 3        & 2           \\ 
\hline
DATING          & 20    & 5         & 25\%   & 0        & 5              & 5            & 0        & 0           \\ 
\hline
EDUCATION       & 417   & 55        & 13.1\% & 0        & 48             & 46           & 9        & 4           \\ 
\hline
ENTERTAINMENT   & 176   & 28        & 15.9\% & 0        & 23             & 23           & 6        & 2           \\ 
\hline
EVENTS          & 9     & 2         & 22.2\% & 0        & 2              & 2            & 0        & 0           \\ 
\hline
FINANCE         & 156   & 13        & 8.3\%  & 1        & 12             & 11           & 1        & 1           \\ 
\hline
FOOD            & 93    & 11        & 11.8\% & 0        & 8              & 8            & 2        & 1           \\ 
\hline
GAME            & 641   & 85        & 13.2\% & 2        & 79             & 77           & 6        & 2           \\ 
\hline
HEALTH          & 167   & 34        & 20.3\% & 0        & 33             & 30           & 2        & 2           \\ 
\hline
HOUSE           & 34    & 2         & 5.9\%  & 0        & 2              & 2            & 0        & 0           \\ 
\hline
LIBRARIES       & 8     & 0         & 0\%    & 0        & 0              & 0            & 0        & 0           \\ 
\hline
LIFESTYLE       & 229   & 30        & 13.1\% & 0        & 26             & 26           & 0        & 2           \\ 
\hline
MAPS            & 74    & 7         & 9.4\%  & 0        & 7              & 7            & 0        & 0           \\ 
\hline
MEDICAL         & 59    & 3         & 5.0\%  & 0        & 3              & 3            & 0        & 0           \\ 
\hline
MUSIC           & 217   & 67        & 30.8\% & 0        & 64             & 60           & 7        & 2           \\ 
\hline
NEWS            & 101   & 28        & 27.7\% & 0        & 28             & 28           & 0        & 0           \\ 
\hline
PARENTING       & 7     & 0         & 0\%    & 0        & 0              & 0            & 0        & 0           \\ 
\hline
PERSONALIZATION & 135   & 18        & 13.3\% & 0        & 17             & 16           & 1        & 1           \\ 
\hline
PHOTOGRAPHY     & 81    & 12        & 14.8\% & 0        & 12             & 12           & 0        & 0           \\ 
\hline
PRODUCTIVITY    & 198   & 23        & 11.6\% & 0        & 17             & 17           & 5        & 3           \\ 
\hline
SHOPPING        & 125   & 15        & 12.0\% & 0        & 14             & 14           & 1        & 1           \\ 
\hline
SOCIAL          & 104   & 25        & 24.0\% & 0        & 23             & 23           & 1        & 1           \\ 
\hline
SPORTS          & 106   & 15        & 14.1\% & 1        & 15             & 14           & 0        & 0           \\ 
\hline
TOOLS           & 283   & 22        & 7.7\%  & 1        & 16             & 15           & 6        & 3           \\ 
\hline
TRAVEL          & 151   & 10        & 6.6\%  & 0        & 10             & 10           & 0        & 0           \\ 
\hline
VIDEO           & 33    & 6         & 18.1\% & 0        & 6              & 6            & 0        & 0           \\ 
\hline
WEATHER         & 20    & 1         & 5.0\%  & 0        & 1              & 1            & 0        & 0           \\
\hline
\end{tabular}
\end{table*}

\subsection{Evaluation Results}
In total we developed 10 algorithms for detecting potential violation of human values in 6 Android APIs. We evaluated our algorithms on our truthset consisting of 46 apps from the Google Play store. Our algorithms detected potential violation of human values with high accuracy and recall performance as shown in \autoref{table:experiment result}.

\begin{table}[H]
\scriptsize
  \centering
\centering
\caption{evaluation results of algorithms}
\label{table:experiment result}
\begin{center}\setlength{\tabcolsep}{1.5mm}{
\begin{tabular}{|l|l|l|l|} 
\hline
\textbf{API name}          & \textbf{violation} & \textbf{Accuracy} & \textbf{Recall}  \\ 
\hline
animation                  & hedonism           & 1.0               & 1.0              \\ 
\hline
\multirow{3}{*}{media}     & hedonism           & 0.761             & 0.808            \\ 
\cline{2-4}
                           & self direction     & 0.761             & 0.808            \\ 
\cline{2-4}
                           & Universalism       & 0.913             & 0.818            \\ 
\hline
mtp                        & self direction     & 1.0               & 1.0              \\ 
\hline
\multirow{2}{*}{nfc}       & security           & 1.0               & 1.0              \\ 
\cline{2-4}
                           & self direction     & 1.0               & 1.0              \\ 
\hline
\multirow{2}{*}{telephony} & security           & 0.957             & 0.5              \\ 
\cline{2-4}
                           & Conformity         & 0.957             & 0.5              \\ 
\hline
hardware                   & security           & 1.0               & 1.0              \\
\hline

\end{tabular}}
\label{tab1}
\end{center}
\end{table}



\begin{table}
\scriptsize
\centering
\caption{Comparison of value violations and security vulnerabilities}
\label{table:virusesandviolation}
\begin{tabular}{|l|l|l|p{1cm}|p{1.5cm}|} 
\hline
Human value                     & API name  & \textcolor[rgb]{0.2,0.2,0.2}{VirusTotal} & Values violation & Overlapping rate  \\ 
\hline
\multirow{3}{*}{Self Direction} & mtp       & 0                                        & 1                & 0.0               \\ 
\cline{2-5}
                                & media     & 157                                      & 975              & 16.10\%           \\ 
\cline{2-5}
                                & nfc       & 2                                        & 14               & 14.29\%           \\ 
\hline
\multirow{3}{*}{Security}       & hardware  & 0                                        & 0                & NA                \\ 
\cline{2-5}
                                & telephony & 74                                       & 117              & 63.25\%           \\ 
\cline{2-5}
                                & nfc       & 34                                       & 53               & 64.15\%           \\ 
\hline
\multirow{2}{*}{hedonism}       & media     & 3                                        & 15               & 20.0\%            \\ 
\cline{2-5}
                                & animation & 0                                        & 0                & NA                \\ 
\hline
Universalism                    & media     & 155                                      & 962              & 16.11\%           \\ 
\hline
Conformity                      & telephony & 74                                       & 117              & 63.25\%           \\
\hline
\end{tabular}
\end{table}


\begin{table}
\scriptsize
\centering
\caption{User comments with number of values violation by APIs}
\label{table:star}
\begin{tabular}{|l|l|l|l|l|l|l|l|} 
\hline
                                &           & \multicolumn{6}{c|}{App comment score}                         \\ 
\hline
Human value                     & API name  & 0   & 0-1                & 1-2                & 2-3 & 3-4 & 4-5  \\ 
\hline
\multirow{3}{*}{Self Direction} & mtp       & 0   & 0                  & 0                  & 0   & 0   & 0    \\ 
\cline{2-8}
                                & media     & 120 & 0                  & 0                  & 11  & 86  & 280  \\ 
\cline{2-8}
                                & nfc       & 5   & .0                 & 0                  & 0   & 2   & 4    \\ 
\hline
\multirow{3}{*}{Security}       & hardware  & 0   & 0                  & 0                  & 0   & 0   & 0    \\ 
\cline{2-8}
                                & telephony & 10  & 0                  & 0                  & 1   & 7   & 14   \\ 
\cline{2-8}
                                & nfc       & 14  & 0                  & 0                  & 1   & 4   & 10   \\ 
\hline
\multirow{2}{*}{hedonism}       & media     & 0   & 0                  & 0                  & 0   & 4   & 1    \\ 
\cline{2-8}
                                & animation & 0   & 0                  & 0                  & 0   & 0   & 0    \\ 
\hline
Universalism                    & media     & 120 & 0                  & 0                  & 11  & 82  & 279  \\ 
\hline
Conformity                      & telephony & 10  & \textasciitilde{}0 & 0\textasciitilde{} & 0   & 7   & 14   \\
\hline
\end{tabular}
\end{table}

\section{Large Scale Analysis of Android Apps}
Based on the promising evaluation results, we carried out a large scale analysis of Android apps using our algorithms to understand potential violation of human values in these different apps and their categories.

\subsection{Dataset}

We randomly selected 10,000 APK files from AndroZoo for our large scale analysis.
AndroZoo  is a very rich APK dataset collected from Google Play by Allix et al., AppChina and Anzhi \cite{b11}. In addition to the APK files, we also collected other information from the Google Play store including user comments, and number of installations.


In our analysis, we consider the root APIs and their sub-APIs and even APIs encapsulated by third-party libraries, e.g., the Google Video Ads API extends the Media API. While some generic root APIs such as android.app and android.animation are widely used by developers, some functional APIs such as android.mtp, android.gesture only exist in an APP that has specific functions, so are used less frequently.

\subsection{Results and Discussion}
Below we highlight and discuss the results of our Android apps analysis.

\subsubsection{Relationship Between Violation of Values and Malware}
VirusTotal\footnote{https://www.virustotal.com/gui/} is a free analysis service for viruses, worms, Trojans, and all kinds of malware that can quickly detect suspicious files and web addresses. The difference between VirusTotal and traditional anti-virus software is that it scans files through various anti-virus engines. Using a variety of anti-virus engines allows users to determine whether an uploaded file is a malicious software based on the detection results of each anti-virus engine. In the AndroZoo dataset, the query results of VirusTotal exist as a feature of the data, so we can measure the performance of our algorithms in terms of security vulnerabilities by comparing value violations with the presence of viruses.



As summarized in \autoref{table:virusesandviolation}, there is a high degree of correlation between viruses and the potential violation of security and conformity values i.e., about 63 per cent of the APK files that contain potential values violation of security and conformity also have viruses. In addition, the degree of correlation between other violations and viruses is minimal in the other value categories. 
This shows that the violation of specific human values in Android apps, such as security, is often accompanied by security vulnerabilities. 

\begin{table*}
\scriptsize
\centering
\caption{App installation frequency with number of values violation by APIs}
\label{table:Installation}
\begin{tabular}{|l|l|l|l|l|l|l|l|l|} 
\hline
                                &           & \multicolumn{7}{c|}{App installation frequency}                                                  \\ 
\hline
Human value                     & API name  & 0 - 100 & 100 - 1000 & 1000 - 10000 & 10000 - 50000 & 50000 - 1e+05 & 1e+05 - 1e+06 & higer 1e+06  \\ 
\hline
\multirow{3}{*}{Self Direction} & mtp       & 0       & 0          & 0            & 0             & 0             & 0             & 0            \\ 
\cline{2-9}
                                & media     & 117     & 86         & 74           & 22            & 56            & 79            & 63           \\ 
\cline{2-9}
                                & nfc       & 4       & .3         & 2            & 2             & 0             & 0             & 0            \\ 
\hline
\multirow{3}{*}{Security}       & hardware  & 0       & 0          & 0            & 0             & 0             & 0             & 0            \\ 
\cline{2-9}
                                & telephony & 7       & 5          & 9            & 6             & 3             & 3             & 2            \\ 
\cline{2-9}
                                & nfc       & 13      & 8          & 5            & 2             & 0             & 1             & 0            \\ 
\hline
\multirow{2}{*}{hedonism}       & media     & 0       & 0          & 2            & 0             & 3             & 0             & 0            \\ 
\cline{2-9}
                                & animation & 0       & 0          & 0            & 0             & 0             & 0             & 0            \\ 
\hline
Universalism                    & media     & 117     & 86         & 72           & 22            & 53            & 79            & 63           \\ 
\hline
Conformity                      & telephony & 10      & 7          & 9            & 3             & 3             & 3             & 2            \\
\hline
\end{tabular}
\end{table*}

\subsubsection{User Perception}

Users may show concern towards violation of their values, e.g., when there are too many security issues in an app or cluttered ads that interfere with the user experience. These concerns are reflected in the reviews and downloads of the app. Hence, we also evaluated additional information of the apps including comments and frequency of installation. \autoref{table:star} and \autoref{table:Installation} summarize user comments and frequency of app installation under different type of violations. In order to  explore the perceptions of users in the comments, we used the overall star rating of an app to represent users' overall satisfaction of the app. Since the lowest star a user can assign to an app is 1 in the Google Play store, in places where the user score was 0, the app did not receive any feedback whatsoever, i.e., there is no star rating or user comment.



As shown in \autoref{table:Installation}, in most APIs, apps with the lowest number of installations had more potential violation of human values. Violations such as NFC and telephony are often found in apps with 50,000 or fewer downloads. A possible explanation is that developers would fix these violations when an app is popular because of profits, reputation, and their responsibility to improve the user experience \cite{b12}. However, in the case of media, the delivery of video ads mainly leads to the violation of the value of pleasure. While advertising can constitute a violation of value because it is such an effective way to monetize, developers usually will not turn down this revenue stream, no matter the size of the downloads.





\subsubsection{App Categories}

Different types of apps may have different emphasis on their functionalities, which leads them to use of different APIs. For example,  \autoref{table:violation frequency for categories} summarizes the total number of different types of apps and corresponding values violation. Even though different apps make use different API services, this does not mean that value violation frequency of apps necessarily relates to the frequency of their API use, e.g., although financial and business apps use many APIs, the value violation ratio for high-security-requirement software remains relatively low. 


Furthermore, social apps such as music, and dating apps show high rates of values violation, i.e., 25\% to 30\%. It can be noted that video advertising makes up the majority of violations. Because of the free nature of these apps, advertising is often the biggest source of revenue \cite{b12}. Although video ads can be highly profitable, it also significantly reduces the user experience.

\section*{Conclusion}
In this paper, we explored the relationships between human values and Android API services based on the literature. We presented algorithms to detect potential violation of human values in 6 API services. The algorithms were evaluated using a manually annotated set of Android apps. Based on the high performance of the algorithms, we carried out a large scale analysis of 10,000 apps covering multiple categories. The results of our analysis show that there is a correlation between the violation of human values and the presence of viruses in apps; apps with the lowest number of installations had more potential of value violations; and social apps had the highest rate of values violation.

Our future work entails conducting empirical studies to understand in finer details how end user values are violated by software artefacts from the perspectives of end users and also the awareness and perception of developers in dealing with these kind of challenges.


\section*{Acknowledgment}
Support for this work from ARC Laureate Program FL190100035 is gratefully acknowledged.


\bibliographystyle{IEEEtran}

\bibliography{reference}

\begin{thebibliography}{10}
\providecommand{\url}[1]{#1}
\csname url@samestyle\endcsname
\providecommand{\newblock}{\relax}
\providecommand{\bibinfo}[2]{#2}
\providecommand{\BIBentrySTDinterwordspacing}{\spaceskip=0pt\relax}
\providecommand{\BIBentryALTinterwordstretchfactor}{4}
\providecommand{\BIBentryALTinterwordspacing}{\spaceskip=\fontdimen2\font plus
\BIBentryALTinterwordstretchfactor\fontdimen3\font minus
  \fontdimen4\font\relax}
\providecommand{\BIBforeignlanguage}[2]{{%
\expandafter\ifx\csname l@#1\endcsname\relax
\typeout{** WARNING: IEEEtran.bst: No hyphenation pattern has been}%
\typeout{** loaded for the language `#1'. Using the pattern for}%
\typeout{** the default language instead.}%
\else
\language=\csname l@#1\endcsname
\fi
#2}}
\providecommand{\BIBdecl}{\relax}
\BIBdecl

\bibitem{b2}
R.~K. Bellamy, K.~Dey, M.~Hind, S.~C. Hoffman, S.~Houde, K.~Kannan, P.~Lohia,
  S.~Mehta, A.~Mojsilovic, S.~Nagar, K.~N. Ramamurthy, J.~Richards, D.~Saha,
  P.~Sattigeri, M.~Singh, K.~R. Varshney, and Y.~Zhang, ``Think your artificial
  intelligence software is fair? think again,'' \emph{IEEE Software}, vol.~36,
  no.~4, pp. 76--80, 2019.

\bibitem{Bloomberg:2021}
S.~Soper, ``Fired by bot at amazon: ‘it’s you against the machine’,''
  [Online.] Available:
  https://www.bloomberg.com/news/features/2021-06-28/fired-by-bot-amazon-turns-to-machine-managers-and-workers-are-losing-out?sref=EJ3iffSv,
  accessed June, 2021.

\bibitem{b8}
S.~H. Schwartz, J.~Cieciuch, M.~Vecchione, E.~Davidov, R.~Fischer,
  C.~Beierlein, A.~Ramos, M.~Verkasalo, J.-E. Lönnqvist, K.~Demirutku,
  O.~Dirilen-Gumus, and M.~Konty, ``\BIBforeignlanguage{eng}{Refining the
  theory of basic individual values},'' \emph{\BIBforeignlanguage{eng}{Journal
  of personality and social psychology}}, vol. 103, no.~4, pp. 663--688, 2012.

\bibitem{Perera:2020}
H.~Perera, A.~Nurwidyantoro, W.~Hussain, D.~Mougouei, J.~Whittle, R.~Shams, and
  G.~Oliver, ``A study on the prevalence of human values in software
  engineering publications, 2015 – 2018,'' in \emph{ICSE}, 2020.

\bibitem{b1}
D.~Mougouei, ``Engineering human values in software through value
  programming,'' p. 133–136, 2020.

\bibitem{b11}
K.~Allix, T.~F. Bissyandé, J.~Klein, and Y.~L. Traon, ``Androzoo: Collecting
  millions of android apps for the research community,'' in \emph{IEEE/ACM 13th
  Working Conference on Mining Software Repositories}, 2016, pp. 468--471.

\bibitem{Cheng:2010}
A.-S. Cheng and K.~R. Fleischmann, ``Developing a meta-inventory of human
  values,'' in \emph{ASIS\&T}, vol.~47, 2010.

\bibitem{Winter:2018}
E.~Winter, S.~Forshaw, and M.~A. Ferrario, ``Measuring human values in software
  engineering,'' in \emph{ACM/IEEE 12th International Symposium on Empirical
  Software Engineering and Measurement}, 2018, pp. 1--4.

\bibitem{Shams:2020}
R.~A. Shams, W.~Hussain, G.~Oliver, A.~Nurwidyantoro, H.~Perera, and
  J.~Whittle, ``Society-oriented applications development: Investigating users'
  values from bangladeshi agriculture mobile applications,'' 2020.

\bibitem{Obie:2020}
H.~O. Obie, W.~Hussain, X.~Xia, J.~Grundy, L.~Li, B.~Turhan, J.~Whittle, and
  M.~Shahin, ``A first look at human values-violation in app reviews,'' in
  \emph{2021 IEEE/ACM 43rd International Conference on Software Engineering:
  Software Engineering in Society (ICSE-SEIS)}, 2021, pp. 29--38.

\bibitem{b3}
M.~Naseri, N.~Borges, A.~Zeller, and R.~Rouvoy, ``Accessileaks: Investigating
  privacy leaks exposed by the android accessibility service,''
  \emph{Proceedings on Privacy Enhancing Technologies}, vol. 2019, pp. 291 --
  305, 2019.

\bibitem{b10}
W.~Wang and M.~W. Godfrey, ``Detecting api usage obstacles: A study of ios and
  android developer questions,'' in \emph{2013 10th Working Conference on
  Mining Software Repositories (MSR)}, 2013, pp. 61--64.

\bibitem{b12}
J.~Bühler, A.~Baur, M.~Bick, and J.~Shi, ``Big data, big opportunities:
  Revenue sources of social media services besides advertising,'' in \emph{14th
  Conference on e-Business, e-Services and e-Society (I3E)}, 2015, pp.
  183--199.

\end{thebibliography}

\end{document}